\documentclass[fleqn,usenatbib]{mnras}

\usepackage{newtxtext,newtxmath}

\usepackage[T1]{fontenc}
\usepackage{lineno}

\DeclareRobustCommand{\VAN}[3]{#2}
\let\VANthebibliography\thebibliography
\def\thebibliography{\DeclareRobustCommand{\VAN}[3]{##3}\VANthebibliography}

\usepackage{graphicx}	
\usepackage{amsmath}	
\usepackage{ulem}

\title[HD 26172: RS CVn binary]{HD 26172: an active solar-type subgiant in a close binary system}

\author[F.-B. Meng et al.]{
	\parbox{\textwidth}{Fang-Bin Meng,$^{1,2}$
		Li-Ying Zhu,$^{1,2}$\thanks{E-mail: zhuly@ynao.ac.cn}
		Sheng-Bang Qian,$^{3}$
		Nian-Ping Liu,$^{1}$
		Jia Zhang,$^{1}$
		David Mkrtichian,$^{4}$
		Soonthornthum Boonrucksar$^{4}$
		and Er-Gang Zhao$^{1}$}
	\vspace{0.5em} \\ 
	$^{1}$Yunnan Observatories, Chinese Academy of Sciences (CAS), 650216, Kunming, People's Republic of China\\
	$^{2}$University of Chinese Academy of Sciences, No.1 Yanqihu East Rd, Huairou District, 101408, Beijing, People's Republic of China\\
	$^{2}$Department of Astronomy, School of Physics and Astronomy, Yunnan University, Kunming 650091, People’s Republic of China\\
	$^{4}$National Astronomical Research Institute of Thailand, 191 Siriphanich Bldg., Huay Kaew Rd., Chiang Mai 50200, Thailand
}

\date{Accepted XXX. Received YYY; in original form ZZZ}

\pubyear{\the\year{}}

\begin{document}
	\label{firstpage}
	\pagerange{\pageref{firstpage}--\pageref{lastpage}}
	\maketitle
	
	\begin{abstract}
		We present the first comprehensive photometric and spectroscopic analysis of the RS CVn system HD 26172, robustly determining the previously debated evolutionary state of its primary star.
		Since this system is a single-lined spectroscopic binary with spot-induced light curve modulations, we derived its physical parameters by combining the TESS light curves, the radial velocity curve from our observations, and the primary-star mass estimates based on three complementary methods.Our results reveal that HD 26172 is a detached binary system composed of a $1.25 \pm 0.32 M_{\odot}$ subgiant and a $0.63 \pm 0.11 M_{\odot}$ main-sequence star. The conclusion of subgiant primary is also supported by the absence of lithium absorption and no observed infrared excess. Using long-term photometry from the KWS survey, we detected a tentative stellar activity cycle of 5635 days with an amplitude of 0.04 mag in HD 26172. Additionally, we identified ten optical flare events exhibiting temporally clustered outburst behavior. The presence of a long-term activity cycle, pronounced starspot activity, and frequent optical flares makes HD 26172 a valuable laboratory for studying magnetic activity in subgiants within close binary systems.
	\end{abstract}
	
	\begin{keywords}
		stars: activity–binaries: eclipsing–stars: individual: HD 26172.
	\end{keywords}
	
	
	
	\section{Introduction}
	RS Canum Venaticorum (RS CVn) binary systems represent an important class of chromospherically active binary stars, first systematically defined by \citet{1976ASSL...60..287H} as close, detached binaries consisting of a subgiant or giant component of spectral type G or K and a late-type main-sequence or subgiant companion, with orbital periods between 1 and 14 days.
	The short orbital periods lead to tidally locked synchronous rotation, driving rapid stellar rotation which in turn promotes magnetic field generation and activity. Key observational signatures of RS CVn systems include asymmetric light curves caused by rotational modulation from large cool spots \citep{1986A&A...165..135R}, enhanced Ca II H$\&$K and H$\alpha$ line emissions \citep{1976ASSL...60..287H}, and luminous X-ray/radio coronae indicative of energetic plasma heating \citep{1980ApJ...236..212W,1985AJ.....90..493M}. Flares spanning optical to X-ray regimes, with energies exceeding solar events by several magnitudes, further highlight their extreme activity \citep{1991ARA&A..29..275H}. In recent years, high-precision and nearly continuous photometric data from satellites such as the Kepler spacecraft and the Transiting Exoplanet Survey Satellite (TESS) have significantly advanced the study of RS CVn systems. These observations not only facilitate the detection of stellar flares but also provide valuable insights into the spatiotemporal evolution of starspots, differential rotation characteristics, and activity cycles, offering crucial observational constraints for understanding stellar magnetic activity \citep{2012A&A...543A.146F,2013ApJ...767...60R,2013MNRAS.432.1203M,2022MNRAS.512.4835M,2023ApJ...948....9I}.

	This paper is concerned with the eclipsing binary HD 26172, which exhibits EA-type light variations \citep{2018ApJS..235....5Q}. Based on ROSAT X-ray observations, HD 26172 exhibits chromospheric activity with $L_X / L_{\text {bol }} \approx 3.5 \times 10^{-4}$, typical of active RS CVn binaries and about two orders of magnitude higher than the Sun $(\sim 10^{-6})$.
	From their joint analysis of spectral energy distributions (SEDs) and light-curve data, \citet{2021ApJ...912..123J} derived for HD\,26172 a radius ratio of 0.32, a semimajor-axis ratio of 3.84, an orbital inclination of $89^\circ$, a TESS-band light ratio of 0.02, and effective temperatures of 6045\,K and 4216\,K for the primary and secondary, respectively. 
	\citet{2023ApJS..264...17Z} discovered excess emission lines in HD 26172 from its LAMOST spectrum. 
	Some previous studies classified HD 26172 as a pre-main-sequence or zero-age main-sequence star based on its location and general properties \citep{1997A&AS..124..449M,1998A&AS..132..173L,2004ChJAA...4..258L}. However, the absence of lithium absorption lines, which are typical in pre-main-sequence stars, and uncertainties in early distance estimates make these classifications inconclusive. Therefore, the evolutionary status of HD 26172 remains poorly constrained, motivating a more detailed investigation in this work.

	The paper is organized as follows. In Section \ref{section 2}, we describe the TESS observations and investigate the relationship between photometric asymmetry and eclipse timing variations. Section \ref{section 3} presents our spectroscopic observations and analysis. In Section \ref{section 4}, we estimate the mass of the primary star and model the light curves from different TESS sectors. Finally, in Section \ref{section 5}, we summarize our findings and discuss the system’s evolutionary status and flare activity statistics.

	\section{Photometric observations and light curve variations}\label{section 2}
	\subsection{Photometric Observations}\label{section 2.1}
\begin{figure*}
		\includegraphics[width=\textwidth]{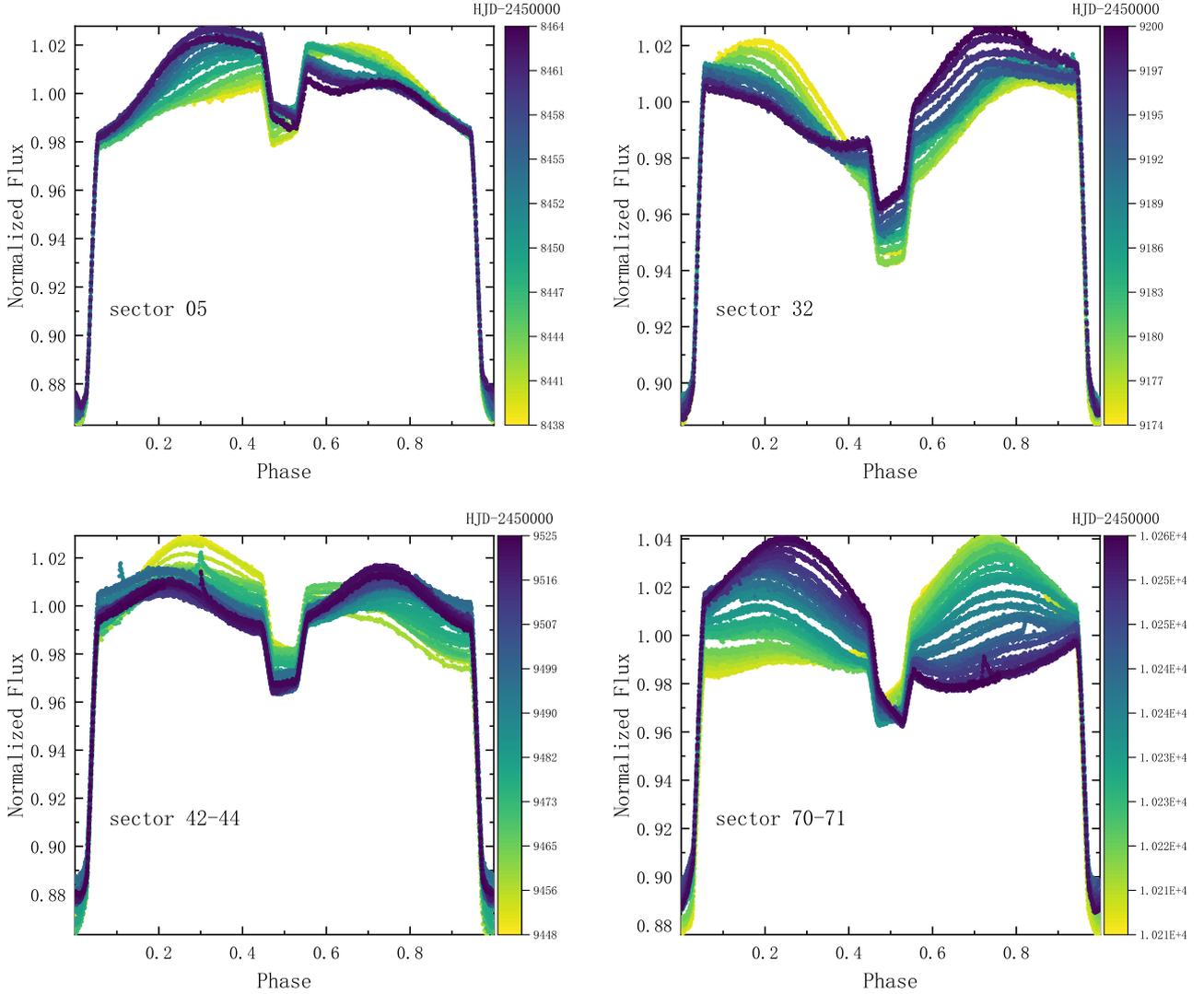}
		\caption{
			TESS light curves of HD 26172 observed across seven sectors, shown on a common time scale (HJD - 2450000). Approximate observing dates are: Sector 05 - Nov–Dec 2018; Sector 32 - Nov–Dec 2020; Sectors 42–44 - Aug–Nov 2021; and Sectors 70–71 - Sep–Nov 2023. Out-of-eclipse variability and occasional flare events are apparent (see Section \ref{section 2.1} for data description).
		}
		\label{LC}
	\end{figure*}

	TESS performed high-precision photometric observations of HD 26172 across seven sectors: 05, 32, 42, 43, 44, 70, and 71. Each sector was observed continuously for approximately 27.4 days \citep{2015JATIS...1a4003R}. The full observational period spanned from HJD 2458438 to 2460260, encompassing the entire range of the seven sectors. The data are publicly available through the Mikulski Archive for Space Telescopes (MAST) and can also be accessed via the Python \texttt{lightkurve} package \citep{2018ascl.soft12013L}. The photometric precision of TESS varies with the target star’s Tmag, with a typical precision of approximately 230 ppm for stars with Tmag = 10, which is much smaller than the amplitude of the observed variability and can be considered negligible for the present analysis. For this analysis, the 2-minute cadence Simple Aperture Photometry (SAP) data were selected. The phase-folded light curves are shown in Figure \ref{LC}. It is clear that continuous luminosity variations are present both in the out-of-eclipse phase and at the minima. Additionally, flares were observed in Sector 43-44 and Sector 70-71. Therefore, the observed variability is likely caused by stellar activity.

	\subsection{Timing Variations Associated with the O’Connell Effect}\label{section 2.2}
	
\begin{figure*}
		\includegraphics[width=\textwidth]{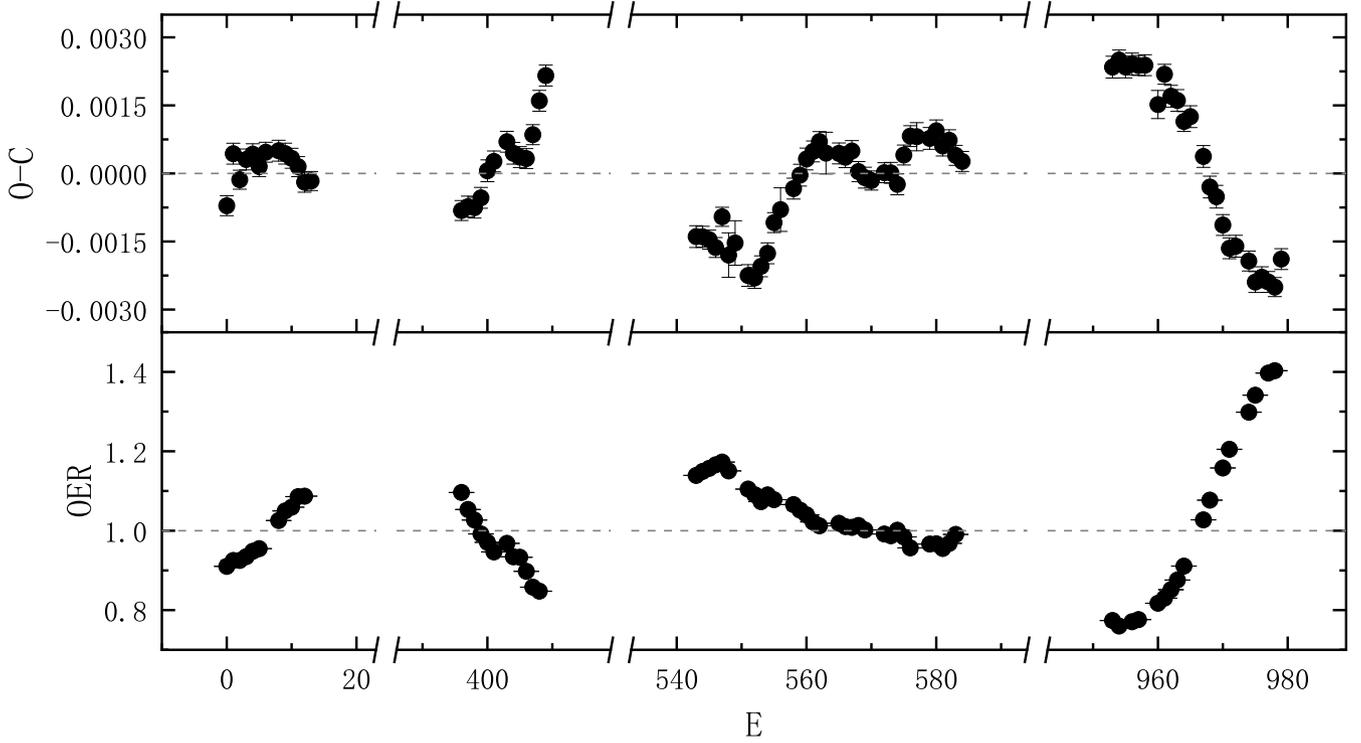}
		\caption{Top panel: O-C diagram after linear correction, with the revised orbital period of 1.8592535(4) days.
			Bottom panel: Variation of the O’Connell Effect Ratio (OER; quantifying the asymmetry of the light curve) with cycle number $E$ (see Section \ref{section 2.2}).} 
		\label{OC-OER}
	\end{figure*}

\begin{figure}
		\includegraphics[width=\columnwidth]{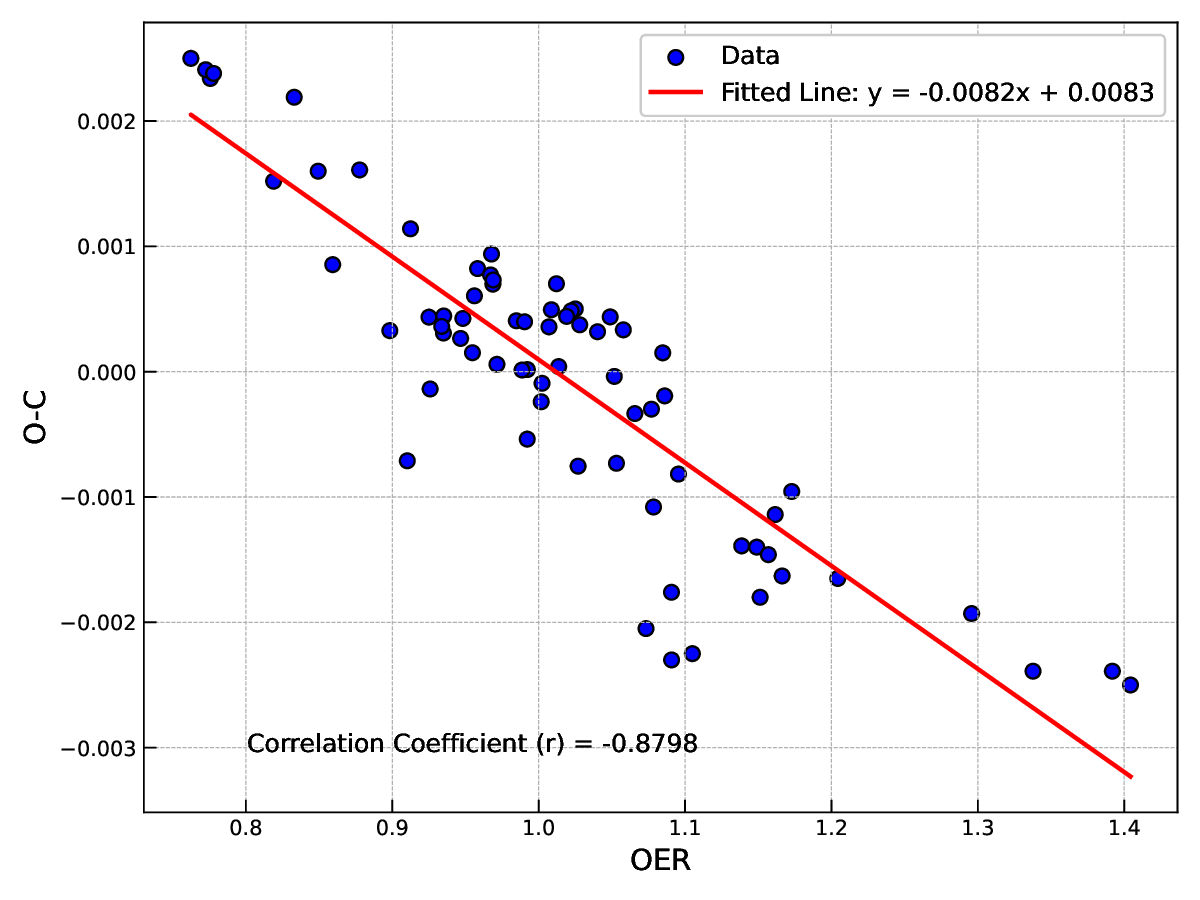}
		\caption{O-C in phase as a function of the O’Connell Effect Ratio (OER; quantifying the asymmetry of the light curve) for all points shown in Fig. \ref{OC-OER}. The scatter plot illustrates the relationship between OER and the corresponding O-C values. The red line represents the fitted linear model, and the Pearson correlation coefficient is $r = -0.8798$  (see Section \ref{section 2.2}) .}
		\label{correlation}
	\end{figure}

	We first analyzed the eclipse timing variations of HD 26172 using all TESS data. Small-amplitude variations are apparent, as shown in the top panel of Figure \ref{OC-OER}. To calculate the eclipse times, we applied a 5-term Fourier series fitting. After linear correction, the period of HD 26172 was revised to 1.8592535(4) days.  The light curve exhibits two maxima of unequal height, a phenomenon known as the O’Connell effect \citep{1951PRCO....2...85O}. To quantify this asymmetry, we calculated the O’Connell Effect Ratio (OER), a metric introduced by \citet{1999PhDT........38M}:
	
	\begin{equation}
		\mathrm{OER}=\frac{\sum_{i=1}^{n / 2} I_i-I_0}{\sum_{i=(n / 2)+1}^n I_i-I_0}
	\end{equation}
	where $n$ is the number of phase bins, $I_i$ is the mean intensity of the $i$-th bin, and $I_0$ is the mean intensity near the light curve minimum. The variations of OER are illustrated in Figure \ref{OC-OER}. For OER and O-C, their trends are opposites. As shown in Figure \ref{correlation}, a linear regression fit of the OER and O-C relationship yields the equation $O-C = -0.0082 \times OER + 0.0083$. The Pearson correlation coefficient (r) is $r=-0.8456$, indicating a strong negative correlation between OER and O-C. As the light curve becomes more asymmetric, the O-C deviates further from zero. This suggests that the variation in eclipse timings is caused by the displacement of the minimum position due to starspots rather than a true change in the orbital period \citep{2013ApJ...774...81T}.

	\section{Spectral observations and analysis} \label{section 3}
	\subsection{Spectroscopic Observations} \label{section 3.1}
	Spectral observations of HD 26172 were conducted with the Beijing Faint Object Spectrograph and Camera \citep[BFOSC;][]{2016PASP..128k5005F} mounted on the 2.16-m telescope at Xinglong Station of the National Astronomical Observatories, Chinese Academy of Sciences. A total of 18 observations were obtained between December 31, 2023, and September 23, 2024, using the E9+G10 configuration with a slit width of 1.6 arcseconds. The instrument covers a wavelength range of 3300–10000 \AA. The spectral resolution per pixel  ranges from 0.38 to 0.86 \AA, corresponding to a resolving power per pixel of 13560–10127. Additionally, medium-resolution spectroscopic observations of HD 26172 were conducted between October 25, 2024, and January 26, 2025, using the 2.4-meter telescope at the Thai National Observatory (TNO), National Astronomical Research Institute of Thailand. During this period, 9 observations were obtained using a medium-resolution spectrograph equipped with an ARC 4K CCD detector. The spectrometer covered a wavelength range from 3800 to 9000 \AA, with a spectral resolution of R $\sim$ 18,000. The Image Reduction and Analysis Facility (IRAF) was used for the processing of the observation images and the extraction of the spectra.
	
	\subsection{Spectral Analysis}\label{section 3.2}
\begin{figure*}
		\includegraphics[width=\textwidth]{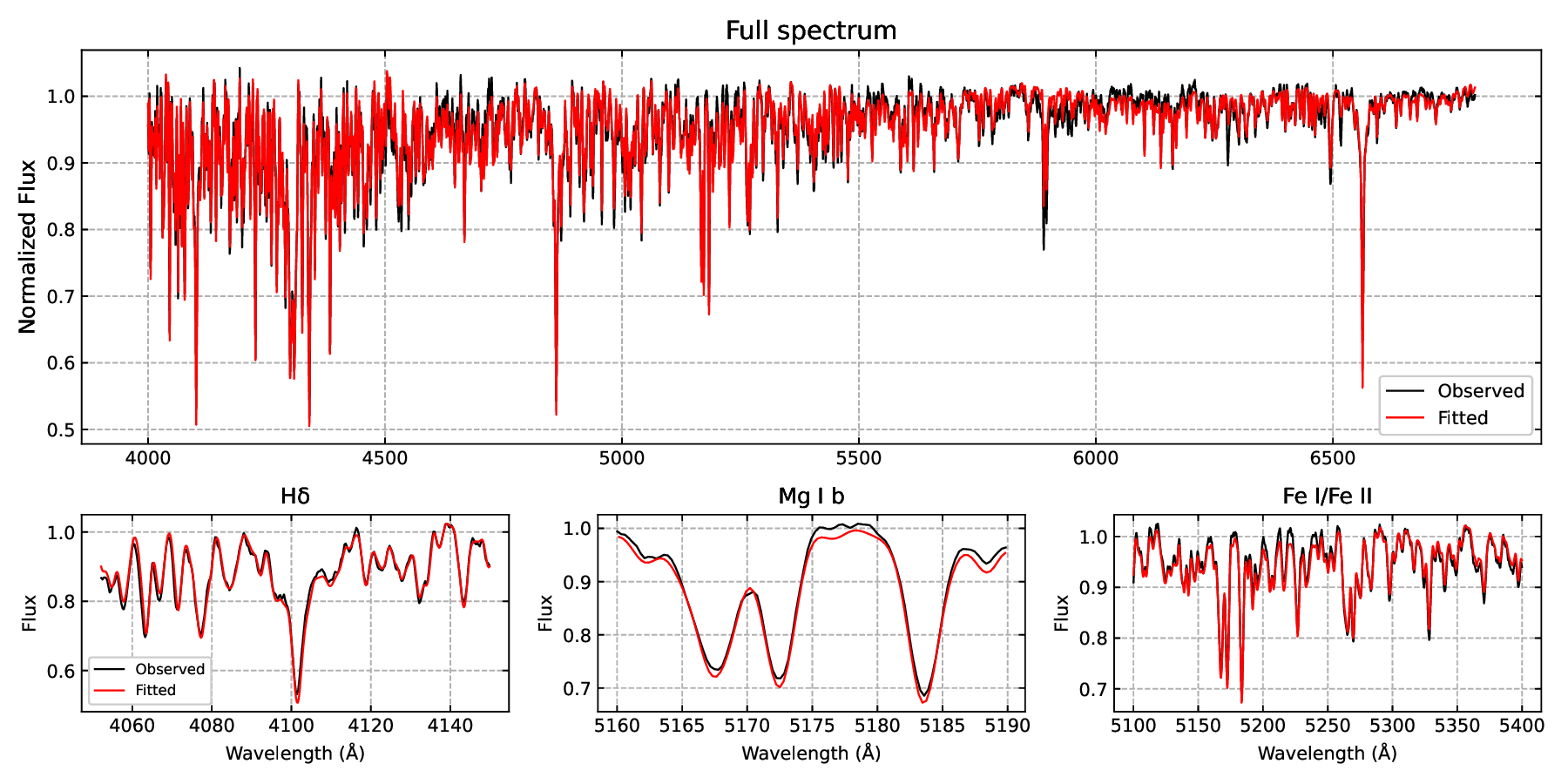}
		\caption{Top panel: fitted spectrum of HD 26172 around phase 0.5, where the secondary star is fully eclipsed and the spectrum represents that of the primary star. The red line denotes the fitted model. Bottom panels: zoomed-in regions sensitive to $T_{\text {eff }}$ and $\log g$ (see Section \ref{section 3.2}).}
		\label{ulyss}
	\end{figure*}

	The stellar atmospheric parameters of HD 26172 were derived by fitting the spectrum obtained around phase 0.5 using the University of Lyon Spectroscopic Analysis Software \citep[ULYSS;][]{2009A&A...501.1269K}. At this phase, the secondary component is fully eclipsed, leaving the spectrum to represent that of the primary component. Spectral fitting was performed using model spectra generated by an interpolator from the ELODIE library\citep{2001A&A...369.1048P}. One of the fitted spectra is shown in Figure \ref{ulyss}  with a red line. The top panel displays the full spectrum, while the bottom panels show three zoomed-in spectral regions particularly sensitive to the effective temperature and surface gravity. The mean values of the stellar atmospheric parameters obtained from this fitting are: $T_{\mathrm{eff}, 1}=5850 \pm 28 \mathrm{~K}, \log g,_1=3.82 \pm 0.09 \mathrm{~g} / \mathrm{cm} \mathrm{~s}^{-2},[\mathrm{Fe} / \mathrm{H}]_1=-0.06 \pm 0.09$ dex.

\begin{figure*}
		\centering
		\includegraphics[width=0.33\textwidth]{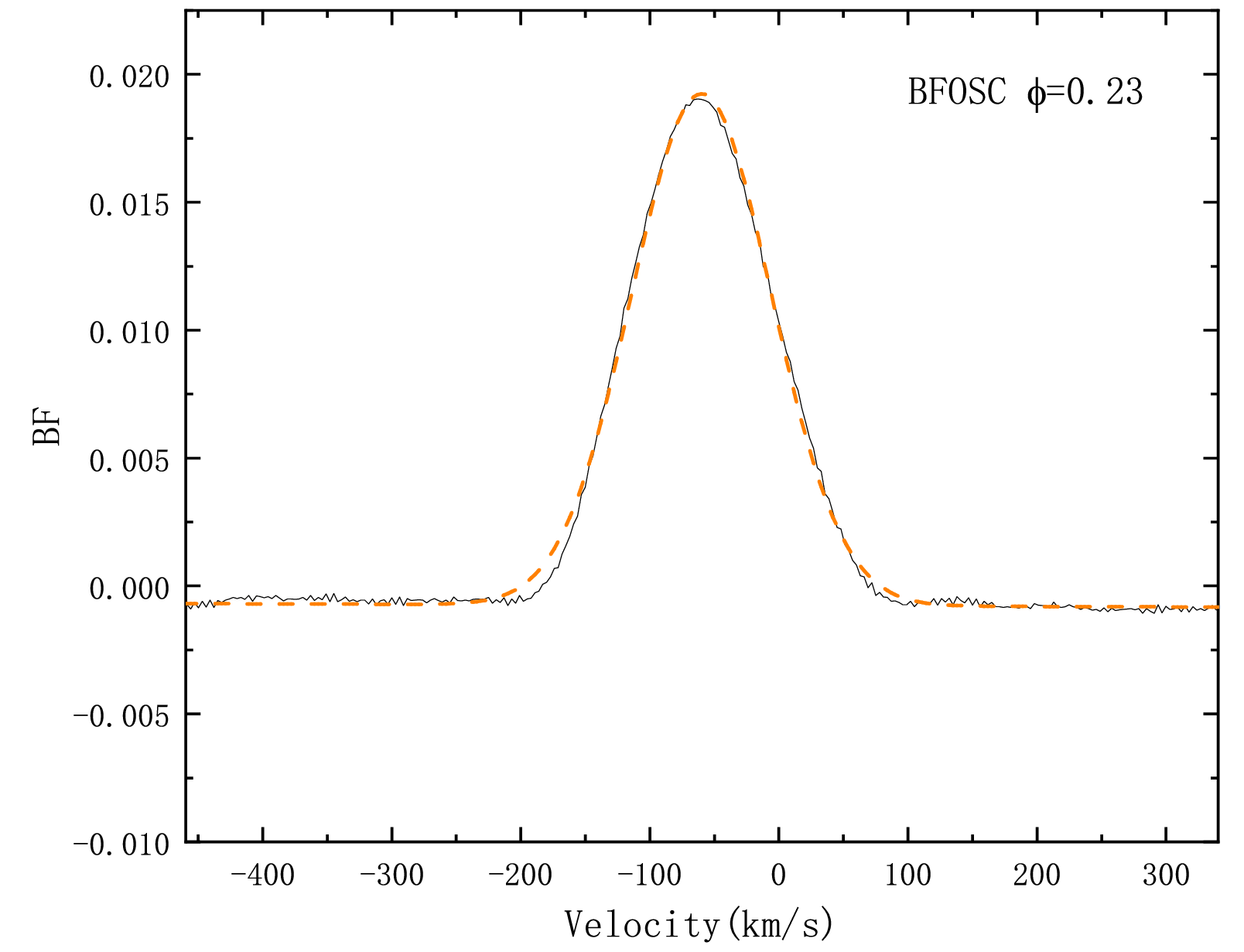}
		\includegraphics[width=0.33\textwidth]{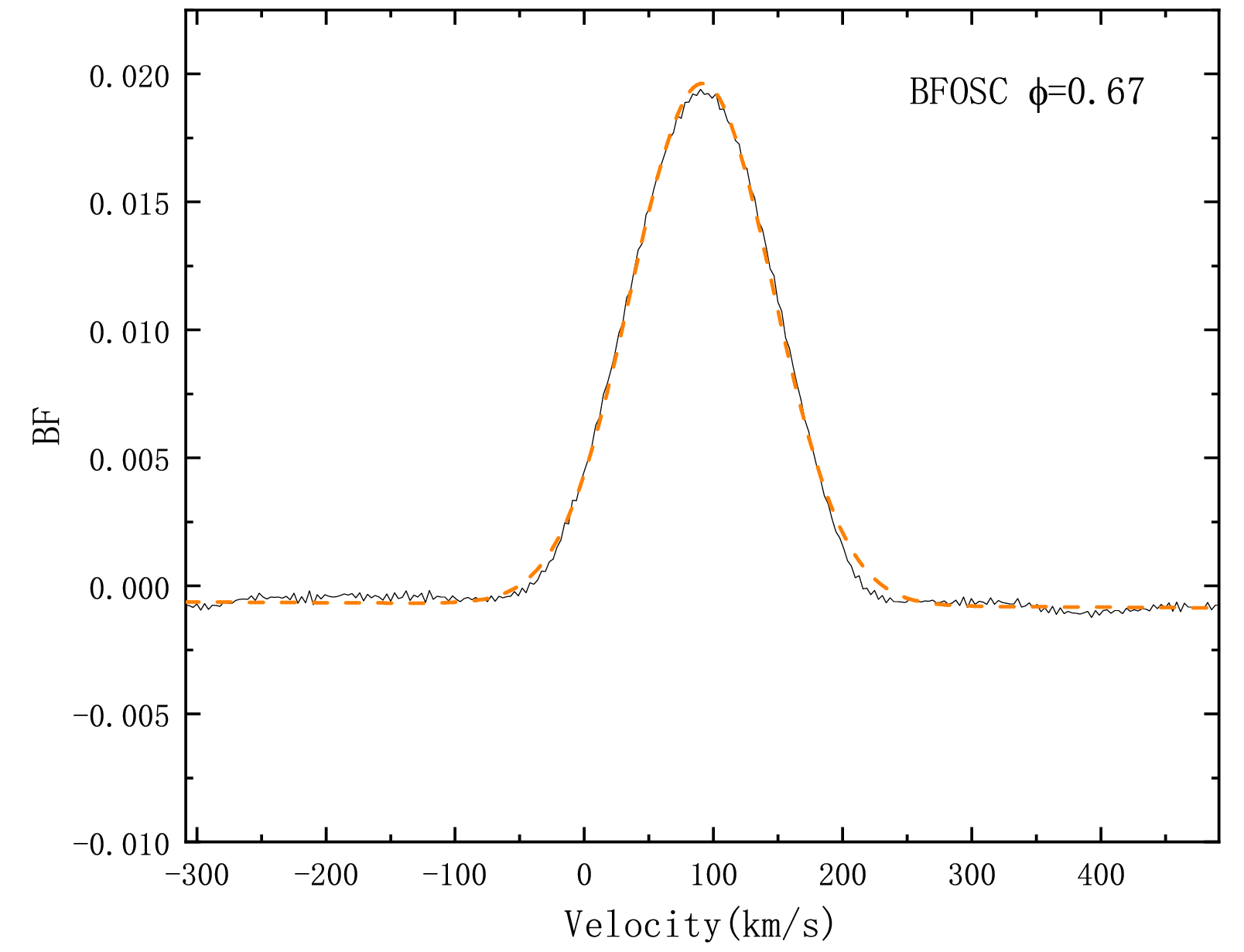}
		\includegraphics[width=0.33\textwidth]{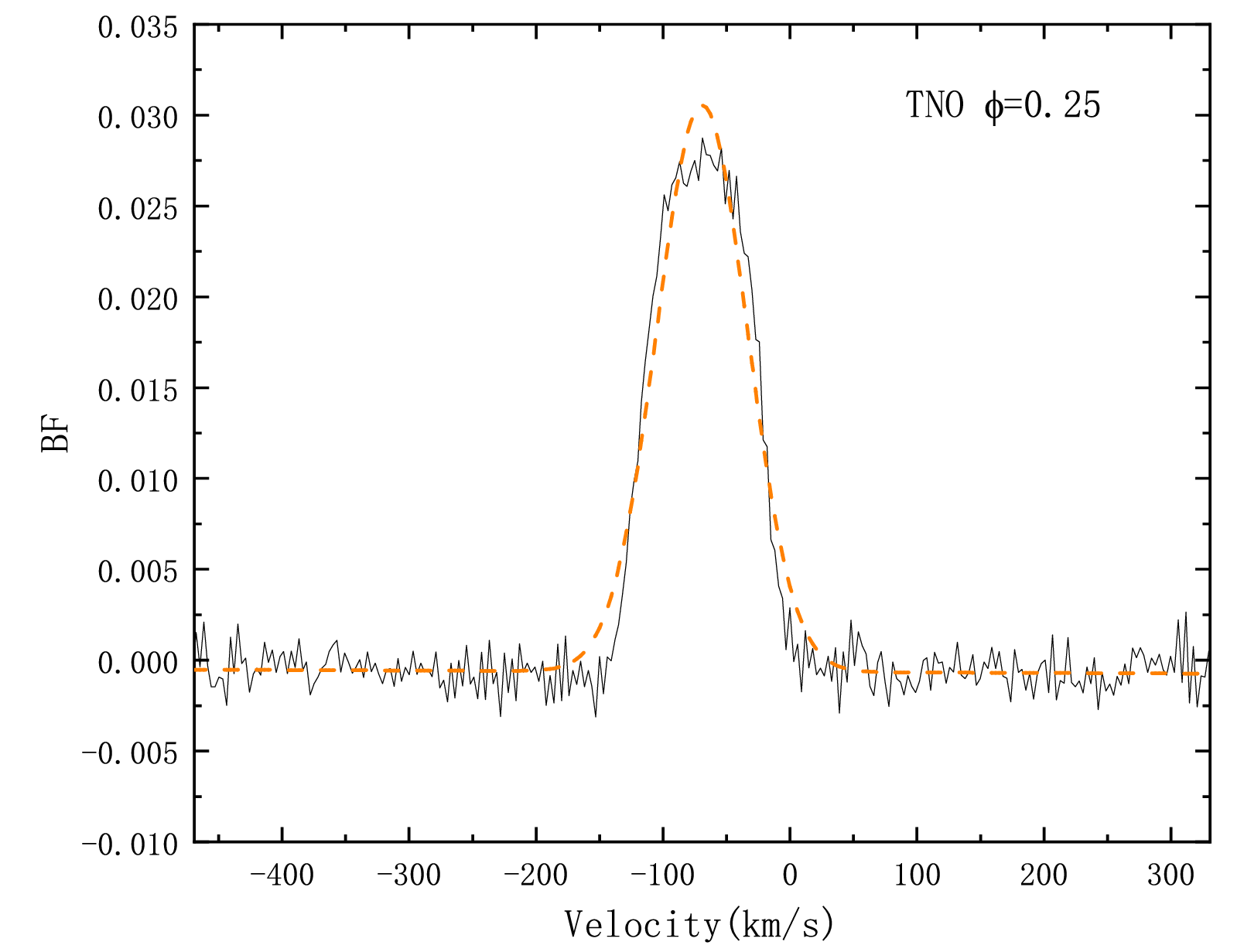}\\
		\caption{Broadening functions (BFs) of HD 26172 obtained at three different orbital phases. Each BF shows a single, well-defined peak with no sign of secondary components, confirming that the spectra are single-lined (see Section \ref{section 3.2}).}
		\label{BF}
	\end{figure*}

\begin{figure}
		\includegraphics[width=\columnwidth]{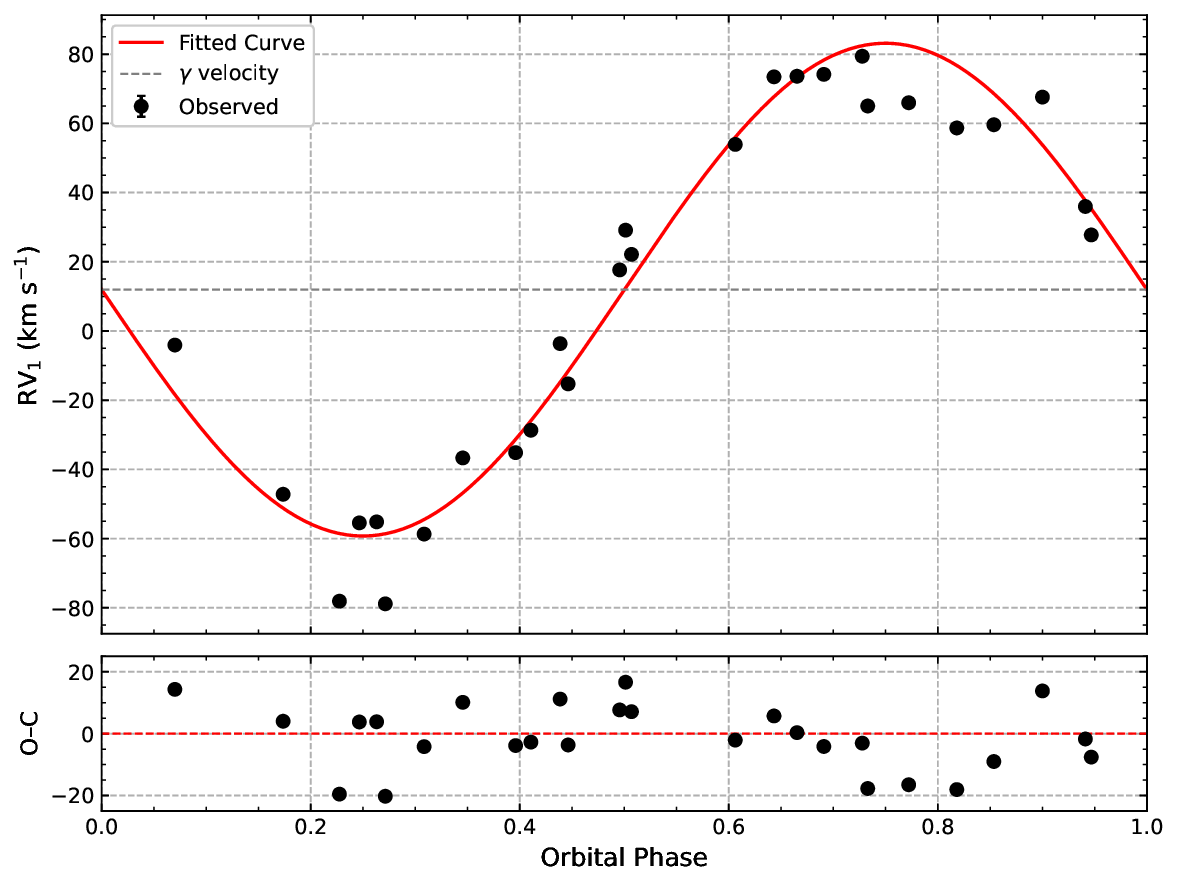}
		\caption{Primary radial velocity curve of HD 26172. Black points show the observed RV data (Table \ref{table_RV}) for one complete orbital cycle (phase 0–1). The red curve represents the fitted sine function (see Section \ref{section 3.2}).}
		\label{RV}
	\end{figure}

	Radial velocities (RVs) were determined from 27 spectra using the broadening function (BF) technique developed by \citet{1999TJPh...23..271R, 2002AJ....124.1746R}. The template spectrum used for extracting BFs was selected from the PHOENIX theoretical spectral library \citep{2013A&A...553A...6H}, with parameters closely matching the target star. RVs were measured by fitting Gaussian functions to the BFs' profiles and calculating the peak positions. Figure \ref{BF} presents three BFs measured at different orbital phases. Each displays a single, well-defined peak with no evidence of secondary components, confirming that the spectra are single-lined. The measured RVs are presented in Table \ref{table_RV}. The RVs were fitted using a sine function of the form 
	$RV = \gamma + K \cdot \sin\left(2 \pi \cdot \left(f \cdot t + \phi \right)\right)$, as shown in Figure \ref{RV}. The fitting yielded a systemic velocity $\gamma = 11.95 \pm 0.11$ km/s and an amplitude $K = 71.20 \pm  2.77$ km/s. The RV uncertainties listed in Table \ref{table_RV} are the formal errors from the Gaussian fitting of the BF and do not represent the actual measurement errors. As inferred from the residuals of the RV curve fitting, the true uncertainties are roughly an order of magnitude larger. The relatively broad shape of the BF is primarily caused by the moderate spectral resolution, which contributes to the larger RV uncertainties.

	\begin{table} 
		\centering
		\caption{Radial-velocity measurements of the primary star in HD 26172 (see Section \ref{section 3.2}).}
		
		\begin{tabular}{ccccc}
			\hline
			\hline
			HJD-245000	&	Phase	&	EXPTIME	&	RV1	&	Error	\\
			(d)	&		&	(s)	&	(km/s)	&	(km/s)	\\
			\hline
		10310.00276 	&	0.667 	&	900	&	73.62 	&	0.70 	\\
		10310.05049 	&	0.693 	&	900	&	74.16 	&	0.74 	\\
		10310.11883 	&	0.729 	&	900	&	79.40 	&	0.73 	\\
		10311.04777 	&	0.229 	&	900	&	-78.08 	&	0.48 	\\
		10311.11390 	&	0.265 	&	900	&	-55.17 	&	0.50 	\\
		10311.98708 	&	0.735 	&	900	&	65.02 	&	0.67 	\\
		10312.05995 	&	0.774 	&	900	&	65.96 	&	0.68 	\\
		10312.14544 	&	0.820 	&	1200	&	58.67 	&	0.70 	\\
		10312.21133 	&	0.855 	&	1200	&	59.60 	&	0.68 	\\
		10312.98788 	&	0.273 	&	900	&	-78.85 	&	0.55 	\\
		10313.05687 	&	0.310 	&	900	&	-58.69 	&	0.51 	\\
		10313.12561 	&	0.347 	&	1200	&	-36.69 	&	0.53 	\\
		10575.35482 	&	0.440 	&	900	&	-3.65 	&	0.58 	\\
		10576.28874 	&	0.943 	&	900	&	35.98 	&	0.63 	\\
		10576.29924 	&	0.949 	&	900	&	27.75 	&	0.62 	\\
		10577.31927 	&	0.497 	&	900	&	17.64 	&	0.64 	\\
		10577.32976 	&	0.503 	&	900	&	29.14 	&	0.61 	\\
		10577.34027 	&	0.509 	&	900	&	22.14 	&	0.69 	\\
		10609.18907 	&	0.645 	&	600	&	73.45 	&	0.48 	\\
		10610.17481 	&	0.175 	&	600	&	-47.19 	&	0.58 	\\
		10610.31031 	&	0.248 	&	600	&	-55.46 	&	0.49 	\\
		10694.22315 	&	0.398 	&	600	&	-35.16 	&	0.62 	\\
		10694.25034 	&	0.412 	&	600	&	-28.68 	&	0.60 	\\
		10695.15988 	&	0.902 	&	600	&	67.59 	&	0.45 	\\
		10696.17504 	&	0.448 	&	600	&	-15.29 	&	0.44 	\\
		10701.05149 	&	0.072 	&	600	&	-4.06 	&	0.38 	\\
		10702.04789 	&	0.608 	&	600	&	53.91 	&	0.51 	\\
			\hline
		\end{tabular}
		\label{table_RV}
	\end{table}

	\section{Absolute Parameters}\label{section 4}
	\subsection{Mass Determination}\label{section 4.1}
	The mass ratio ($q$) plays a crucial role in determining the geometric configuration and physical properties of eclipsing binaries. However, for detached systems, the light-curve morphology is largely insensitive to $q$ \citep{2005Ap&SS.296..221T}, and the single-lined nature of HD 26172, together with strong starspot modulation, makes the conventional "q-search" method unreliable. To obtain a reliable estimate of $q$, we first estimated the primary mass using three complementary methods, and then combined these estimates with the spectroscopic mass function to constrain the plausible range of $q$.

	\subsubsection{Isochrone-Based Estimate}\label{section 4.1.1}
	We first estimate the evolutionary mass of the primary star using stellar isochrones from the PARSEC database \citep{2012MNRAS.427..127B}, following the procedure described by \citet{2019ApJS..244...43Z}. The observed atmospheric parameters ($T_{\rm eff}$, $\log g$, [Fe/H]) were compared with the isochrone grid within a three-sigma parameter space, defined as $T_{\rm eff} \pm 3\sigma_{T_{\rm eff}}$, $\log g \pm 3\sigma_{\log g}$, and [Fe/H] $\pm 3\sigma_{\rm [Fe/H]}$. All grid points satisfying these conditions were included in the matching process. Among them, the best-fitting point—minimizing the distance in the ($T_{\rm eff}$, $\log g$, [Fe/H]) space—was adopted as the representative solution. The uncertainties in the derived parameters were estimated from the maximum and minimum values of all isochrone points within this three-sigma range, thereby accounting for both observational errors and model dispersion. This procedure yields two plausible evolutionary solutions: a pre-main-sequence (PMS) track and a post-main-sequence (subgiant) track. We adopt the subgiant branch solution, corresponding to  $M_1 = 1.26_{-0.12}^{+0.10},M_\odot$ and $R_1 = 2.26_{-0.29}^{+0.36},R_\odot$, with an estimated age of $\sim4.2$~Gyr. The alternative PMS interpretation and its implications are discussed in Section \ref{Evolutionary State}.

	\subsubsection{SED-Based Estimate}\label{section 4.1.2}
\begin{figure*}
		\includegraphics[width=\textwidth]{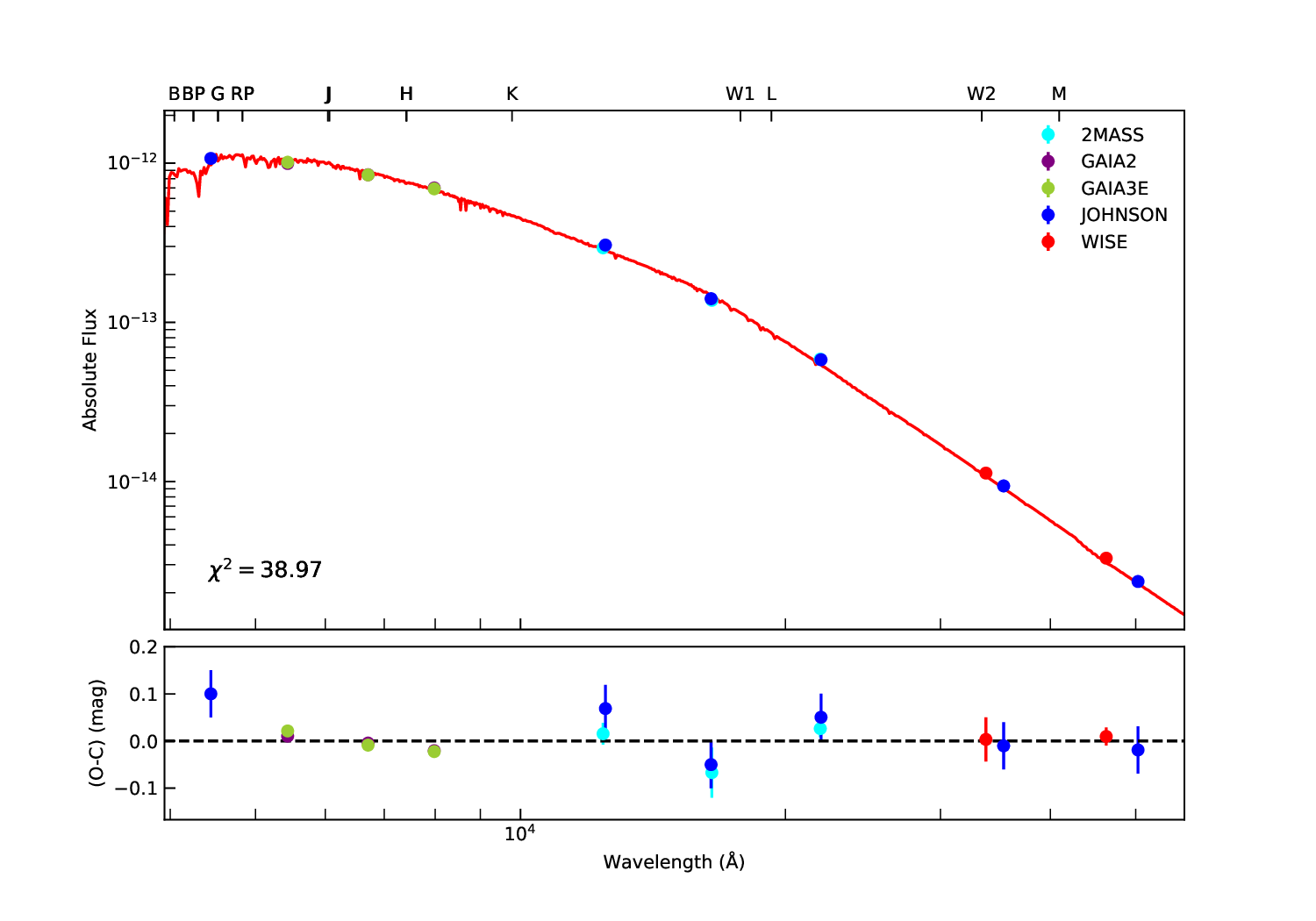}
		\caption{The observed spectral energy distribution (SED; black points) of HD 26172 
			together with the best-fitting SED model (red curve), computed using the 
			speedyfit package (see Section \ref{section 4.1.2}). The fit incorporates multi-band 
			photometry from Gaia, APASS, 2MASS, and WISE.}
		\label{SED}
	\end{figure*}

	Given the large luminosity contrast between the two components—where the secondary contributes only about 2$\%$ of the total light as inferred from light curve modeling—and the absence of secondary features in the BF profiles of our spectra (see Section \ref{section 3.2} and Figure \ref{BF}), we adopted a single-star model for SED fitting. Attempts to use a binary model yielded unphysical parameters for the secondary, further justifying this choice. The fitting was performed using the Python package \texttt{speedyfit}\footnote{\url{https://github.com/vosjo/speedyfit}}
	\citep{2012A&A...548A...6V,2013A&A...559A..54V,2017A&A...605A.109V,2021A&A...655A..43V}, which fits synthetic stellar atmosphere models to multi-band photometric data, including Gaia EDR3 \citep{2021A&A...649A...1G}, APASS \citep{2015AAS...22533616H}, 2MASS \citep{2003yCat.2246....0C}, and WISE \citep{2014yCat.2328....0C}. The best-fit parameters were derived through a Markov Chain Monte Carlo (MCMC) sampling procedure, which also yields robust uncertainty estimates. Gaia parallax and spectroscopic constraints were included to improve reliability. From the SED, the derived surface gravity and radius yield a primary mass of $M_1 = 1.23_{-0.26}^{+0.27} M_{\odot}$. The best-fit result is shown in Figure \ref{SED}.
	
	\subsubsection{Luminosity-Based Estimate}
	Following \citet{2021AJ....162...13L}, we first derived the system’s total luminosity, $L_T=L_1+L_2$, using the apparent magnitude, Gaia parallax, interstellar extinction (from our SED fitting), and bolometric corrections from \citet{2011ApJS..193....1W}. For totally eclipsing systems, the temperature and radius ratios between the two components can be estimated from the eclipse depths and contact timings. The temperature ratio can be estimated from the relative eclipse depths under the assumption of blackbody radiation and negligible limb darkening (standard relation; e.g., \citealp{2018Ap&SS.363...19K}): 
	
	\begin{equation}
		\frac{T_2}{T_1} \approx \left( \frac{d_2}{d_1} \right)^{1/4},
	\end{equation}
	Similarly, the radius ratio can be computed from the geometry of total eclipses using standard expressions for contact timings:
	
	\begin{equation}
		\frac{R_2}{R_1} \approx \frac{t_4 - t_1 - t_3 + t_2}{t_4 - t_1 + t_3 - t_2},
	\end{equation}
	where $d_1$ and $d_2$ are the primary and secondary eclipse depths, and $t_1$ through $t_4$ denote the standard contact points: $t_1$ and $t_4$ mark the beginning and end of the eclipse, while $t_2$ and $t_3$ indicate the start and end of totality, respectively. To accurately measure the eclipse depths and contact timings, we first removed out-of-eclipse variations from the light curve—which may arise from stellar activity—using a LOWESS (Locally Weighted Scatterplot Smoothing) algorithm. The detrended light curve was then phase-folded, and the resulting eclipse profile was used to determine the radius and temperature ratios. With $L_{\mathrm{T}}, R_2 / R_1$, and $T_2 / T_1$, the luminosity of the primary was obtained as
	
	\begin{equation}
		L_1=\frac{L_{\mathrm{T}}}{1+\left(R_2 / R_1\right)^2\left(T_2 / T_1\right)^4}
	\end{equation}
	Adopting the effective temperature estimate from spectral modeling, $T_1 = 5850 \pm 28$ K, we derived the radius $R_1$ using the Stefan–Boltzmann relation. Finally, using the spectroscopic $\log g_1=3.82 \pm 0.09$, we computed the primary mass as $M_1=1.31 \pm 0.29 M_{\odot}$.
		The absolute parameters derived from the three complementary methods are summarized in Table  \ref{table_para}

	\subsection{Binary Modeling}\label{section 4.2}
	The mass of the primary component, obtained consistently from the three complementary approaches described above, allows for the estimation of the mass ratio. Using the parameters of a single-lined spectroscopic binary and the primary mass, the mass ratio was computed from the standard mass function,
	
	\begin{equation}
		f(m)=\frac{(M_2 \sin i)^3}{\left(M_1+M_2\right)^2}=\left(1.036 \times 10^{-7}\right)\left(1-e^2\right)^{3 / 2} K_1^3 P \mathrm{M}_{\odot}.
	\end{equation}
	By substituting the three primary masses derived from these approaches (see Table \ref{table_para}) into the above equation, three values of $q$ were obtained. Their arithmetic mean, $q = 0.50 \pm 0.05$, was adopted as the final result, since the uncertainties of the three individual mass estimates are roughly similar. The quoted uncertainty ($\pm 0.05$) reflects both the propagated errors and the scatter among the three $q$ values.
 The light curve (see Figure \ref{LC}) shows that the secondary eclipse occurs at phase 0.5, suggesting a circular orbit. Therefore, a null eccentricity ($e=0$) was assumed in the analysis.  The orbital inclination, derived from photometric solutions for different mass ratios, shows little variation.
	
\begin{figure*}
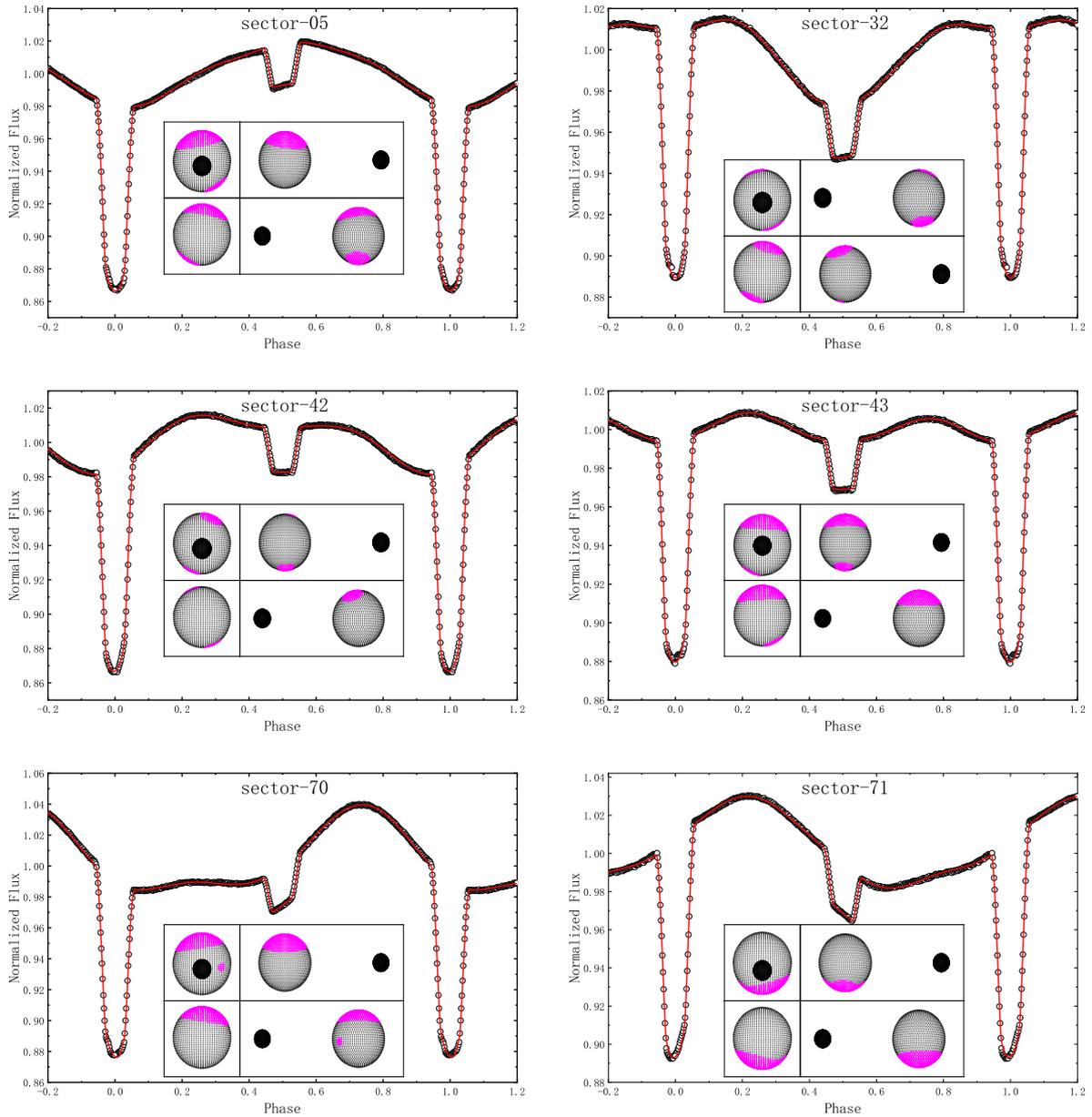

		\centering
		\includegraphics[width=0.45\textwidth]{W-D-S05.eps}
		\includegraphics[width=0.45\textwidth]{W-D-S32.eps}\\
		\includegraphics[width=0.45\textwidth]{W-D-S42.eps}
		\includegraphics[width=0.45\textwidth]{W-D-S43.eps}\\
		\includegraphics[width=0.45\textwidth]{W-D-S70.eps}
		\includegraphics[width=0.45\textwidth]{W-D-S71.eps}
		\caption{Light-curve solutions including starspots, derived from W–D modeling for different TESS sectors. Only six sectors are displayed, and the pink areas mark the fitted spot locations (see Section \ref{section 4.2}). The “north/south” labels of polar spots are model coordinates and do not indicate the true hemisphere, due to the inherent north–south degeneracy of the fit.}
		\label{WD}
	\end{figure*}

	With the mass ratio fixed at $q = 0.50$, the light curves from each TESS sector were analyzed using the Wilson–Devinney (W–D) code \citep{1971ApJ...166..605W,2012AJ....144...73W,2014ApJ...780..151W}.
	The W–D code models the light curves of eclipsing binaries by solving the Roche geometry, taking into account the ellipsoidal effect, gravity darkening, reflection, and limb darkening.
	In our analysis, we fitted the photometric data by iteratively adjusting parameters such as the orbital inclination, mean surface temperature of star 2, surface potentials, bandpass luminosity of star 1, and spot properties until the synthetic and observed light curves converged. Based on the stellar atmospheric parameters derived from spectral fitting (see Section \ref{section 3.2} and Table \ref{table_para}), the effective temperature of the primary component is set to 5850 K. For the convective atmosphere, the gravity darkening coefficients are set as $g_1=g_2=0.32$, and the bolometric albedos are $A_1=A_2=0.5$  \citep{1967ZA.....65...89L,1969AcA....19..245R}. The bolometric and bandpass limb-darkening coefficients were taken from van Hamme \citep{1993AJ....106.2096V}. The light curves of HD 26172 exhibit a clear asymmetry between the two maxima. Such asymmetry is generally attributed to large-scale starspot activity on one of the components, as supported by Doppler imaging studies of RS CVn-type binaries \citep{2005LRSP....2....8B}. Guided by this interpretation, we introduced starspots in our light-curve modeling to reproduce the observed distortions. For each TESS sector, a portion of the light curve covering one complete orbital cycle was selected for spot modeling, to avoid complications from spot evolution over longer timescales. We tested various spot configurations on both components, and found that only spots on the more luminous primary component can account for the observed asymmetries, as expected given its dominant contribution to the system’s light. The modeling started with a single cool spot, and if this could not adequately reproduce the observed light curve, an additional spot was introduced until a satisfactory fit was achieved. The spot temperature factor was fixed during the fitting process to reduce parameter degeneracy, as the spot radius and temperature factor are known to be strongly correlated in the W–D model. The parameters were refined through iterative adjustments to achieve the optimal fit. The fitted models are presented in Figure \ref{WD}, where six sectors are shown as examples (the complete set includes all seven sectors). The resulting models successfully reproduce the overall light-curve morphology and the out-of-eclipse modulation.
	The starspot modeling provides a plausible physical explanation for the observed photometric variability. The orbital parameters derived from the light-curve modeling are listed in Table \ref{table_para}. For each parameter, the value reported corresponds to the mean over the fitted sectors, and the uncertainty is given by the standard deviation among the sectors.

	\begin{table}
		\centering
		\caption{Fundamental Parameters of HD 26172}
		
		\begin{tabular}{cc}
			\hline
			\hline
			Stellar Atmospheric Parameters & \\
			\hline
			$T_{\mathrm{eff}, 1} (K) $  & $5850 \pm 28$\\
			$[\mathrm{Fe} / \mathrm{H}]_1 (dex)$& $-0.06 \pm 0.09$ \\
			$\log g,_1 (\mathrm{g} / \mathrm{cm} \mathrm{~s}^{-2})$ &  $3.82 \pm 0.09$ \\
			\hline
			Isochrone-Based Estimate &  \\
			\hline
			$R_1 (R_{\odot})$ &  $2.26_{-0.29}^{+0.36}$\\
			$M_1 (M_{\odot})$ &  $1.26_{-0.12}^{+0.10}$\\
			$L_1 (L_{\odot})$ &  $5.35_{-1.28}^{+1.80}$\\
			\hline
			SED-Based Estimate &\\
			\hline
			$R_1 (R_{\odot})$ &  $2.32_{-0.04}^{+0.04}$\\
			$M_1 (M_{\odot})$ &  $1.23_{-0.26}^{+0.27}$\\
			$L_1 (L_{\odot})$ &  $5.36_{-0.28}^{+0.30}$\\
			$ebv$& $0.06_{-0.02}^{+0.02}$\\
			\hline
			Luminosity-Based Estimate &\\ 
			\hline
			$m_v $(mag) &  $8.7750 \pm 0.0008$\\
			$parallax$&  $7.26 \pm 0.11$\\
			$BC$&  $-0.08 \pm 0.03$\\
			$T_2/T_1$&  $0.6941 \pm 0.0003$\\
			$R_2/R_1$&  $0.3149 \pm 0.0008$\\
			$R_1 (R_{\odot})$ &  $2.34 \pm 0.08$\\
			$M_1 (M_{\odot})$ &  $1.31\pm0.29$\\
			$L_1 (L_{\odot})$ &  $5.76\pm0.40$\\
			\hline
			Light Curve Modeling &  Detached Model\\
			$q (M_2/M_1)$ &0.50 (fixed) \\
			$i (^{\circ})$ &$87.97 \pm 0.98$ \\
			$T_1$ (K) & 5850 (fixed)  \\
			$T_2$ (K) & $4129 \pm 20$\\
			$\Omega_1$ & $4.31 \pm 0.03$\\
			$\Omega_2$ & $7.29 \pm 0.12$\\
			$f_1 $ & $0.2179 \pm 0.0056$ \\
			$f_2 $ & $0.0173 \pm 0.0010$\\
			$R_1 (R_{\odot})$ &  $2.08 \pm 0.16$\\
			$M_1 (M_{\odot})$ &  $1.25 \pm 0.32$\\
			$L_1 (L_{\odot})$ &  $4.54 \pm 0.70$\\
			$R_2 (R_{\odot})$ &  $0.65 \pm 0.05$\\
			$M_2 (M_{\odot})$ &  $0.63 \pm 0.11$\\
			$L_2 (L_{\odot})$ &  $0.11 \pm 0.02$ \\
			$A (R_{\odot})$ & $7.85 \pm 0.59$\\
			\hline
		\end{tabular}
		\label{table_para}
	\end{table}
	
	\section{Discussions and Summary}\label{section 5}
	
	\subsection{Spots and Flares}\label{section 5.1}
	The light curve of HD 26172 exhibits distinct O'Connell effect variations in each cycle, i.e., the primary component undergoes significant spot reconfigurations over just a few rotations. By comparing these variations with the small-amplitude O-C changes, we find that both arise from the distortion of the light curve caused by dark spots. Modeling of the light curve indicates the presence of large spots near the "poles" of the primary component (in the W–D model, "poles" refer to the geometric latitude 0 = north, $\pi$ = south). Similar high-latitude spots are commonly observed in RS CVn-type binaries through Doppler imaging \citep{1999A&A...347..225S} and are known to be long-lived \citep[e.g.,][]{2016MNRAS.456..314X,2024NatCo..15.9986S}. The spot parameters listed in Table~\ref{tab:spots} represent physically plausible solutions, but they are not necessarily unique.

	\begin{table}
		\centering
		\caption{Spot parameters (in radians) for different TESS sectors.}
		\resizebox{\columnwidth}{!}{
			\begin{tabular}{@{}lcccccc@{}}
				\hline
				\hline
				Sector & Co-Lat$_1$ & Long$_1$ & Radius$_1$ & Co-Lat$_2$ & Long$_2$ & Radius$_2$ \\
				\hline
				Sector 05 & 2.58926 & 1.50092 & 0.45517 & 0.20476 & 5.26741 & 0.90300 \\
				Sector 32 & 2.70404 & 1.95504 & 0.43652 & 0.38063 & 3.85739 & 0.53514 \\
				Sector 42 & 2.74916 & 4.83185 & 0.31251 & 0.45934 & 0.95081 & 0.40186 \\
				Sector 43 & 2.71323 & 4.53256 & 0.35459 & 0.06704 & 1.65689 & 0.99630 \\
				Sector 44 & 2.77225 & 4.23836 & 0.28142 & 0.30201 & 1.73621 & 0.22425 \\
				Sector 70 & 1.63546 & 0.73304 & 0.10792 & 0.15600 & 4.72380 & 0.97252 \\
				Sector 71 & 2.90131 & 1.80211 & 0.95232 & 2.62382 & 4.61172 & 0.43908 \\
				\hline
			\end{tabular}
		}
		\vspace{2mm}
		\noindent\textit{Note.} The temperature factor for all spots was fixed at 0.85 during the modeling.
		\label{tab:spots}
	\end{table}

\begin{figure*}
		\includegraphics[width=\textwidth]{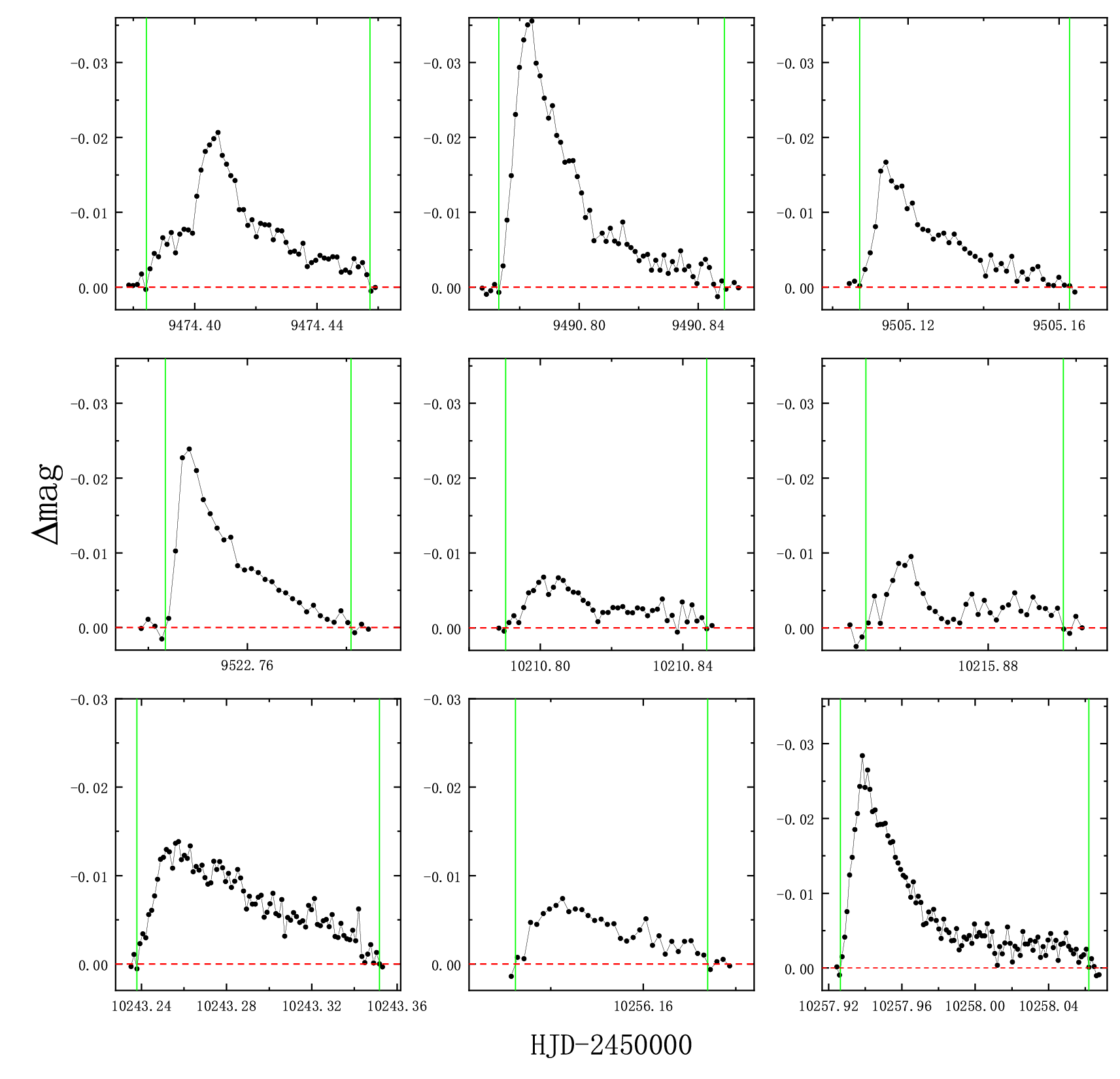}
		\caption{Samples of flare light curves with the eclipsing light curve subtracted. The green solid line indicates the start and end times of the flare, while the red dashed line represents the brightness of the star in a quasi-quiescent state (see Section \ref{section 5.1}).}
		\label{flare}
	\end{figure*}

	After removing the binary photometric variations, a total of ten flares were detected during TESS Sectors 43–44 and 70–71, with amplitudes ranging from 7 to 36 mmag (Figure \ref{flare}). The rise time, decay time, and amplitude of each flare were measured directly from the light curves. The flare energies were then estimated following the method of \citet{2013ApJS..209....5S}, which assumes that a flare radiates approximately as a blackbody with a fixed temperature. The instantaneous bolometric luminosity was derived from the observed flux enhancement and the stellar flux, and the total flare energy was obtained by integrating the flare luminosity over time. For HD 26172, the primary contributes 98$\%$ of the total system flux, so the observed flux variations are dominated by the primary. Using the primary’s radius and effective temperature ensures that the stellar flux matches the observed flux; this does not imply that the flares necessarily occurred on the primary, but provides a reasonable approximation for the flare energy calculation. The typical uncertainty of this approach is about 60$\%$, mainly due to uncertainties in the stellar luminosity and the blackbody assumption. All detected flares occurred within two consecutive TESS observing intervals, while no flares were found in earlier sectors. This temporal clustering suggests that flare activity may have been enhanced during epochs of stronger magnetic activity. Notably, these TESS sectors correspond to times near the brightness maximum of the long-term modulation revealed by ground-based photometry (see Section \ref{section 5.2} and Figure \ref{KWS}), possibly indicating a connection between the occurrence of flares and the phase of the magnetic activity cycle. Nevertheless, given the small number of detected events, this trend remains tentative and requires further observations for confirmation.

	\begin{table} 
		\centering
		\caption{The parameters of the flares detected from the short-cadence data of HD 26172.}
		\begin{tabular}{ccccc}
			\hline
			\hline
			$t_{peak}$	&	$t_{rise}$	&	$t_{decay}$	&	$\Delta M_{max}$	&	$E_{flare}$	\\
			(HJD-2450000)	&	(min)	&	(min)	&	(mmag)	&	(erg)	\\
			\hline
			9501.31947 	&	25 	&	52 	&	8 	&	2.76E+34	\\
			10215.86243 	&	15 	&	50 	&	10 	&	2.76E+34	\\
			10256.14251 	&	15 	&	45 	&	7 	&	2.96E+34	\\
			10210.80089 	&	15 	&	66 	&	7 	&	3.11E+34	\\
			9505.11415 	&	10 	&	70 	&	17 	&	5.74E+34	\\
			9522.74829 	&	7 	&	47 	&	24 	&	5.75E+34	\\
			9474.40765 	&	34 	&	72 	&	21 	&	1.04E+35	\\
			10243.25730 	&	28 	&	136 	&	14 	&	1.48E+35	\\
			9490.78402 	&	16 	&	93 	&	36 	&	1.50E+35	\\
			10257.93839 	&	17 	&	178 	&	28 	&	1.93E+35	\\
			\hline
		\end{tabular}
		\label{table_flare}
	\end{table}

	\subsection{Long-term Magnetic Activity Variations}\label{section 5.2}
	
\begin{figure}
		\includegraphics[width=\columnwidth]{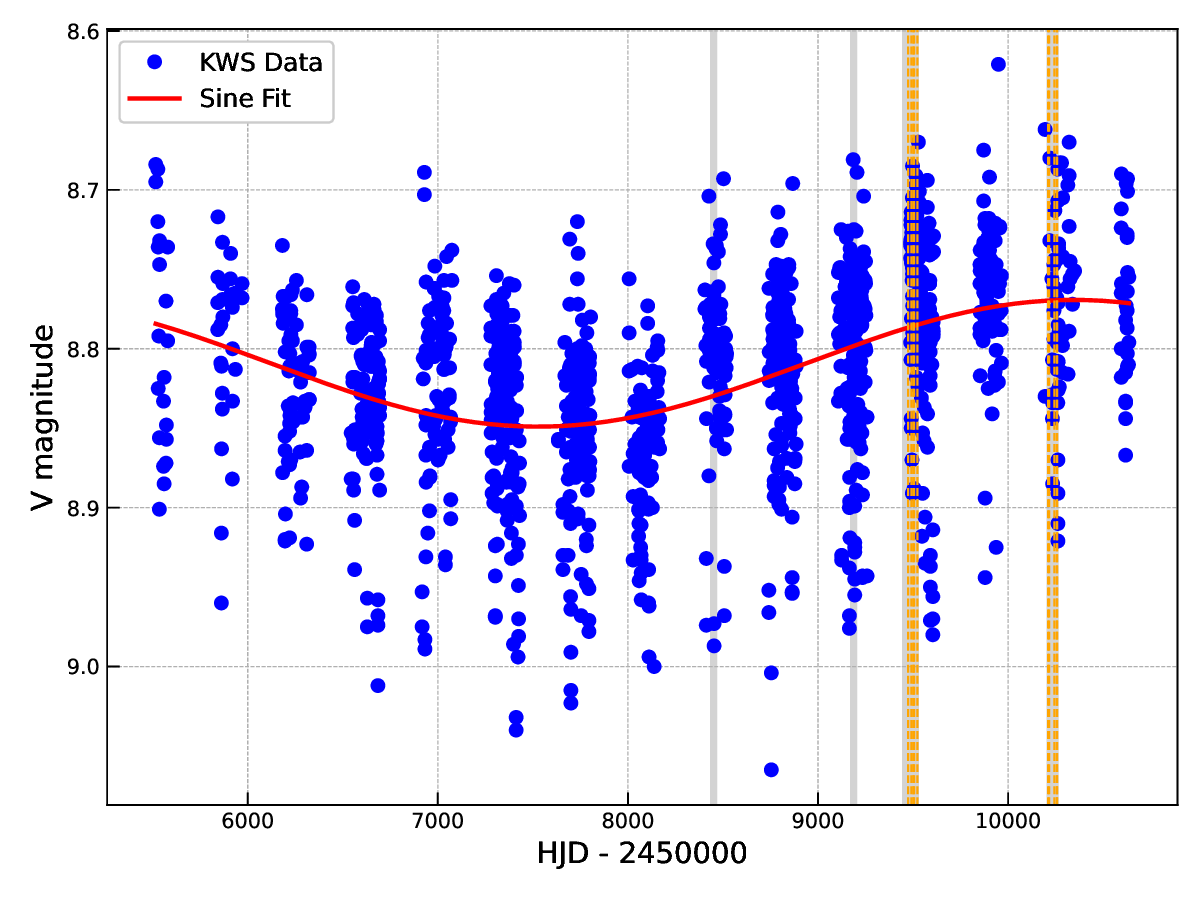}
		\caption{The KWS light curve of HD 26172 showing a quasi-sinusoidal long-term modulation. The curve illustrates a sinusoidal fit used to characterize the long-term variability, while acknowledging that the inferred timescale is comparable to the observational baseline. The orange dashed lines indicate the epochs of detected flares, and the grey shaded areas mark the TESS-covered intervals  (see Section \ref{section 5.1} and \ref{section 5.2}).}
		\label{KWS}
	\end{figure}

	To study the long-term variations potentially associated with stellar magnetic activity,
	we collected observational data from the Kamogata–Kiso–Kyoto Wide-field Survey
	(KWS\footnote{\url{http://kws.cetus-net.org/}}), covering HJD $2455516 - 2460635$.
	We fitted a sine function to the KWS light curve in order to characterize its long-term variability. This yields a characteristic timescale of $P \approx 5635 \pm 366$ days and an amplitude of $A = 0.040 \pm 0.003$ Vmag, although the inferred timescale is comparable to the observational baseline and should therefore be considered as a phenomenological description rather than a confirmed cycle. In Figure~\ref{KWS} the KWS light curve is shown together with the fitted sinusoidal trend. The times of detected flares are marked with semi-transparent orange vertical dashed lines, while the TESS observational coverage for each sector is indicated by shaded grey intervals. Similar long-term brightness modulations are common in RS CVn-type binaries. For example, \citet{2022MNRAS.512.4835M} studied 120 southern systems and claimed activity cycles in 91 of them, based on long-term photometric variability. Such variations are commonly interpreted as arising from stellar dynamos driven by rapid rotation, which induce changes in starspot coverage and modulate the observed brightness.
	
\begin{figure*}
		\centering
		\includegraphics[width=0.45\textwidth]{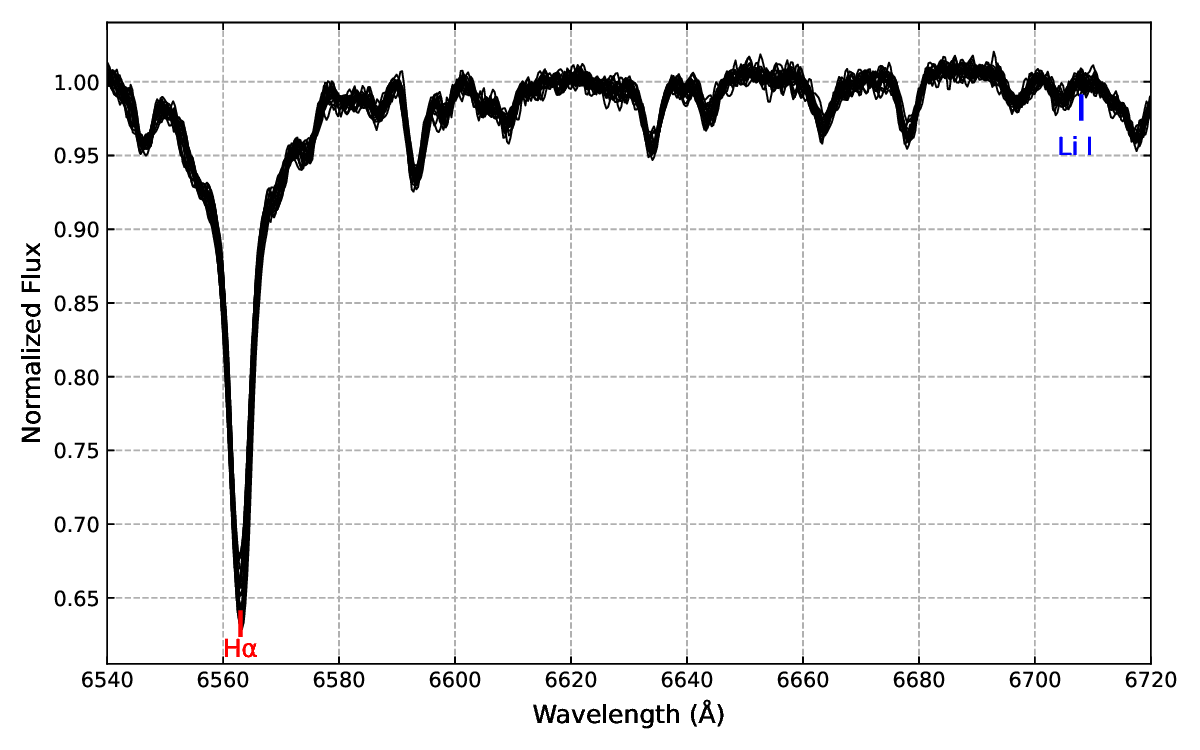}
		\includegraphics[width=0.45\textwidth]{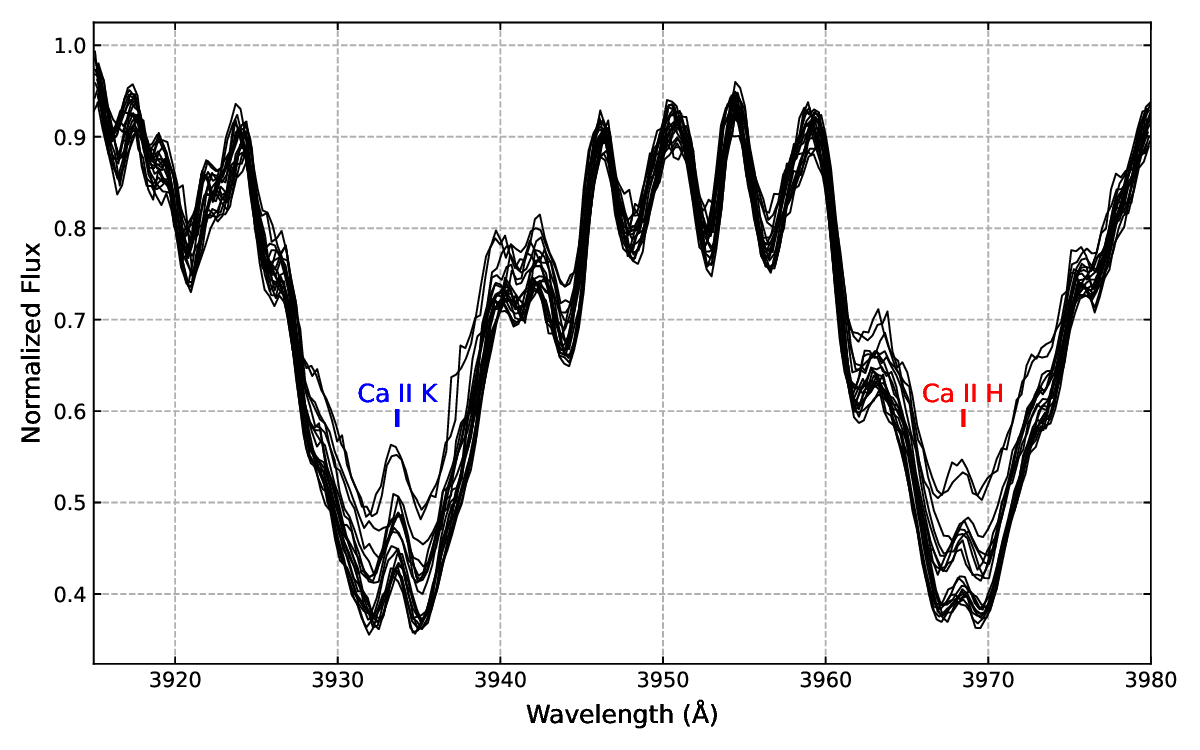}\\
		\caption{The radial velocity-aligned BFOSC spectra of HD 26172. The left panel shows the spectra centered on H$\alpha$ and Li I 6708\,\AA, and the right panel shows the spectra centered on Ca II H$\&$K lines (see Section \ref{section 5.2}).}
		\label{BFOSC}
	\end{figure*}

	In addition to the photometric evidence, our spectroscopic observations further confirm the presence of chromospheric activity in HD 26172. As shown in Figure \ref{BFOSC}, we co-aligned all available BFOSC spectra of HD 26172 in the radial velocity frame. The H$\alpha$ profiles exhibit only slight variations over time. In contrast, the Ca II H$\&$K lines display significant variability across epochs. Both lines consistently show filled-in absorption profiles with distinct core emission features, which are characteristic signatures of chromospheric activity and typically originate from the middle chromosphere \citep{2015RAA....15..252Z}. Such behavior is fully consistent with the known properties of RS CVn-type binaries, which commonly exhibit long-lived magnetic activity manifested through both photospheric spots and chromospheric emission.

	\subsection{Evolutionary State}\label{Evolutionary State}
	HD 26172 is located at RA = 04:08:49.6, Dec = +10:27:49.17, with a distance of approximately 138 pc, known with an uncertainty of about 2 pc $(\approx 1.5 \%)$ according to Gaia DR3. Its sky position and parallax place it within the projected extent of the Taurus–Auriga star-forming region \citep{1997A&AS..124..449M,1998A&AS..132..173L}. However, neither the studies by \citet{1997A&AS..124..449M} and \citet{1998A&AS..132..173L}, nor our own spectroscopic observations (see Figure \ref{BFOSC}), reveal any significant Li I 6708 \AA{} absorption feature. The absence of lithium strongly argues against a PMS nature for HD 26172, as PMS stars normally retain significant lithium abundances.
	
	To further assess its membership status, we employed the BANYAN $\Sigma$ tool \citep{2018ApJ...856...23G}, a Bayesian classification algorithm that identifies likely members of young stellar associations within 150 pc of the Sun using full 6D Galactic kinematics (XYZUVW). Given the Gaia DR3 astrometry and radial velocity of HD 26172, BANYAN $\Sigma$ returns a membership probability of only 13.8$\%$ for the Taurus association, and 86.2$\%$ for the field population, favoring the interpretation that the star is not a current member of any known young moving group.

	From our derived atmospheric parameters, the primary component of HD 26172 may formally lie either on the PMS track or along the subgiant branch. According to \citet{2012MNRAS.427..343M}, the star shows no significant infrared excess, consistent with our SED fitting (see Figure \ref{SED}), which covers the available near- to mid-infrared wavelengths. This indicates the absence of substantial circumstellar material, although faint far-infrared emission cannot be completely ruled out. In addition, comparison of its LAMOST spectrum with those of several subgiants of similar atmospheric parameters (e.g., UCAC4 557-011022 and UCAC4 476-001564) shows that HD 26172’s overall absorption-line pattern is consistent with subgiants and lacks the characteristic features of PMS stars. Taken together, these lines of evidence support the classification of HD 26172 as a subgiant, consistent with its interpretation as a newly identified RS CVn-type binary. Future observations, particularly infrared spectra capable of detecting the radial velocity of the secondary component, would help to independently determine the mass ratio and further constrain the system’s evolutionary status. The W-D modeling indicates that HD 26172 is a detached system, with filling factors of $f_1=$ $0.2178 \pm 0.0056$ for the primary component and $f_2=0.0172 \pm 0.0011$ for the secondary component. As evolution continues, the primary will fill its Roche lobe first and may initiate mass transfer to the secondary component, potentially evolving into a precursor of an Algol-type binary system.

	\subsection{Summary}
	We have presented a comprehensive analysis of the RS CVn-type binary HD 26172 using TESS photometry, KWS long-term data, and spectroscopic observations. The main results can be summarized as follows:
	
	1. Although HD 26172 is a single-lined spectroscopic binary, its absolute parameters were derived through three different methods. The system consists of a subgiant primary and a main-sequence secondary, consistent with typical RS CVn binaries. The evolved primary, with a deeper convective envelope, likely drives the high magnetic activity observed in the system.
	
	2.The W–D modeling reveals a large, polar spot, similar to those found in II Peg \citep{2014MNRAS.438.2307X}, CF Tuc \citep{1999MNRAS.305..966B}, IM Peg \citep{2000A&A...360..272B}, and V711 Tau \citep{1999ApJS..121..547V}. Such polar-dominated structures suggest a long-lived, global magnetic field sustained by an efficient stellar dynamo.
	
	3.During TESS observations, HD 26172 exhibited intense flare activity. The detected flares show an average duration of approximately 99 minutes and an occurrence rate of 0.4$\%$, both of which significantly exceed the median values reported for typical RS CVn-type binaries \citep{2025A&A...699A.322X}, indicating an exceptionally enhanced level of magnetic reconnection activity.
	
	4.The spectra display Ca II H $\&$ K core emission, a hallmark of chromospheric activity in RS CVn binaries, confirming the system’s high activity level.
	
	5.The KWS light curve shows a quasi-sinusoidal modulation of roughly 15 yr, a behavior commonly observed in RS CVn systems and attributed to stellar dynamo activity \citep{2022MNRAS.512.4835M}. The limited temporal coverage makes the period tentative.
	
	Overall, HD 26172 exhibits all the key features of an active RS CVn binary -- spots, chromospheric Ca\,II H\&K core emission, frequent flaring, and long-term brightness modulation -- indicating that it is a highly active member of the RS CVn class.

	\section*{Acknowledgements}
	We gratefully acknowledge the referee for the helpful comments and suggestions, which significantly improved the quality of this work.
	This work was supported by the International Partnership Program of the Chinese Academy of Sciences (Grant No. 020GJHZ2023030GC), the Yunnan Fundamental Research Projects (Grant Nos. 202503AP140013, 202401AS070046, and 202501AS070055), the Yunnan Revitalization Talent Support Program, and the Chinese Academy of Sciences President’s International Fellowship Initiative (Grant No. 2025PVA0089).
	This research has made use of data collected by TESS, which are publicly available from the Mikulski Archive for Space Telescopes (MAST). Funding for the TESS mission is provided by NASA's Science Mission Directorate. We acknowledge the support of the staff of the Xinglong 2.16m telescope. This work was partially supported by National Astronomical Observatories, Chinese Academy of Sciences. We also thank the staff of the 2.4m telescope at the Thai National Observatory (TNO), operated by the National Astronomical Research Institute of Thailand (NARIT), for their assistance during spectroscopic observations. This research has made use of data from the Kyoto University Wide-field Survey (KWS) and from the European Space Agency (ESA) mission Gaia, processed by the Gaia Data Processing and Analysis Consortium (DPAC).
	
	\section*{Data Availability}
	The data underlying this article are available at the Mikulski Archive for Space Telescopes (MAST) (\url{https://mast.stsci.edu/}) and the Kyoto University Wide-field Survey (KWS) (\url{http://kws.cetus-net.org/}) .

	
	
	\bibliographystyle{mnras}
	\bibliography{example} 

	
	


	\bsp	
	\label{lastpage}
\end{document}